# An octave spanning mid-infrared frequency comb generated in a silicon nanophotonic wire waveguide

Bart Kuyken[1,2], Takuro Ideguchi[3], Simon Holzner[3], Ming Yan[3,4], Theodor W. Hänsch[3,4], Joris Van Campenhout[5], Peter Verheyen[5], Stéphane Coen[6], Francois Leo[1,2], Roel Baets[1,2], Gunther Roelkens[1,2], Nathalie Picqué[3,4]

[1]*Photonics Research Group, Department of Information Technology, Ghent University–imec, Ghent, Belgium*
[2]*Center for Nano- and Biophotonics (NB-Photonics), Ghent University, Ghent, Belgium*
[3]*Max Planck Institut für Quantenoptik, Hans-Kopfermann str. 1, 85748 Garching Germany*
[4]*Ludwig-Maximilians-Universität München, Fakultät für Physik, Schellingstr. 4/III, 80799 Munich, Germany*
[5]*imec, Kapeldreef 75, Leuven, Belgium*
[6]*Physics Department, The University of Auckland, Private Bag 92019, Auckland, New Zealand*
*Bart.Kuyken@intec.ugent.be*

**Abstract:** We demonstrate an octave-spanning frequency comb with a spectrum covering wavelengths from 1,540 nm up to 3,200 nm. The supercontinuum is generated by pumping a 1-cm long dispersion engineered silicon wire waveguide by 70 fs pulses with an energy of merely 15 pJ. We confirm the phase coherence of the output spectrum by beating the supercontinuum with narrow bandwidth CW lasers. We show that the experimental results are in agreement with numerical simulations.

## 1. Introduction

Frequency combs in the mid-infrared region [1] have been mostly generated by nonlinear frequency conversion of near-infrared frequency combs. Though the field is currently very active with the exploration of many different and promising approaches [2, 3], producing a very broad spectrum with a slowly varying envelope remains challenging. Supercontinuum generation in a highly nonlinear fiber is known, under certain circumstances [4], to be a powerful way to generate an octave-spanning frequency comb.

However, in the mid-infrared spectral region, suitable materials have remained scarse and difficult to engineer. The only demonstration of a coherent octave-spanning frequency comb spectrum so far has been achieved by spectral broadening of a frequency comb generator based on 3.1-µm optical parametric oscillation of 250 pJ pulse energy [5] in a nonlinear chalcogenide fiber. The difficulty to produce such chalcogenide fibers, the need for tapering them in situ and the breakage of the fiber under high average pump power [5], limit the versatility of the approach. The fragility of the chalcogenide glasses as well as the deterioration of these glasses in the presence of moisture is another major issue. More recently, a hybrid silica-chalcogenide structure, where the chalcogenide is embedded in a silica fiber for protection, has shown to generate a coherent supercontinuum, albeit not octave spanning [6]. However, such devices remain to have a limited lifetime of a few hours to a couple of days depending on the pump powers used [6]. Another approach is the use of quasi-phase matched periodically-poled lithium niobate (PPLN) waveguides. Impressive results have been obtained and an octave spanning phase coherent supercontinuum has been generated [7]. However absorption between 2,500 nm and 2,800 nm and more importantly the limited transparency of lithium niobate beyond 4,500 nm, inhibits the scaling of the technology to longer wavelengths. Furthermore high energy pulses (7 nJ) are needed due to the moderate nonlinearity of the waveguide. Additionally, during the poling of the crystal small random variations on the location of the walls of the poled domains are introduced. This aberration increases the conversion of parasitic processes significantly [7, 8] and makes modeling difficult. Silicon-based waveguides have been originally conceived for the telecommunication region. In this region, octave-spanning supercontinuum generation has been demonstrated by pumping silicon nitride waveguides with 150 pJ pulses centered at 1.3 µm [9], but the coherence conservation in the supercontinuum generation process has not been investigated.

Recently the application of silicon technology to the mid-infrared spectral region has attracted significant interest. Silicon nanophotonic wire waveguides can be engineered [10] within a nanometer precision in a standard CMOS facility. Such waveguides offer many advantages for mid-infrared nonlinear optics, mostly related to the wide transparency range of silicon (1.1 - 8 µm), its high nonlinear refractive index, the possibility of precise dispersion engineering of the waveguide platforms and the high refractive index contrast between the silicon waveguide core and the cladding material (typically $SiO_2$ or air), which allows for densely integrated waveguide systems with a nonlinear parameter an order to two orders of magnitude higher than possible in chalcogenide or silicon nitride systems. In this letter, we report on the design of strongly nonlinear, dispersion controlled silicon photonic wire waveguides. We harness such waveguides for mid-infrared supercontinuum generation and we demonstrate a phase-coherent frequency comb generator with a 30 dB bandwidth spanning from 1,540 nm up to 3,200 nm with coupled input pulse energies as low as 16 pJ.

**2. Device Structure**

The photonic wire is fabricated in a CMOS pilot line [10] on a 200-mm silicon-on-insulator (SOI) wafer and consisting of a 390-nm thick silicon device layer on top of a 2-µm buried oxide layer. The inset in Figure 1a) shows a schematic cross section of the silicon photonic wire. The 1-cm-long air-clad photonic wire has a rectangular cross-section of 1,600 nm x 390 nm. The waveguide is slightly over etched by 10 nm into the buried oxide. The photonic wire widens near the cleaved facets to a 3-µm wide waveguide section for improved coupling efficiency. As a result of the high nonlinear index of silicon [11] and the strong optical confinement obtained by the high linear refractive index of silicon, the nonlinear parameter in the silicon wire is 38 $(Wm)^{-1}$ at 2,300 nm for the highly confined quasi-TE mode. Such high nonlinear parameter in silicon waveguides shows the advantage of using silicon over chalcogenide tapers ($\gamma = 4.5$ $(Wm)^{-1}$ [12]) and silicon nitride waveguides ($\gamma = 1.2$ $(Wm)^{-1}$ [9]) where the nonlinear parameter is much lower. As a result of the high confinement, the waveguide dispersion of the silicon photonic wire contributes strongly to the overall dispersion of the optical waveguide, such that group velocity dispersion can be engineered by optimizing the waveguide dimensions. The group velocity dispersion of the quasi-TE mode of the dispersion engineered photonic wire waveguide as a function of wavelength is shown in Figure 1a). The group velocity dispersion is simulated with the help of a finite element mode solver (Fimmwave). The zero dispersion wavelength is at 2,180 nm and the dispersion becomes positive (normal) at shorter wavelengths, while the dispersion remains low over a wide spectral band. By using a cut-back technique the propagation loss for the quasi-TE mode is determined to be < 0.2 $dB.cm^{-1}$ in the wavelength range of 2,200-2,400 nm.

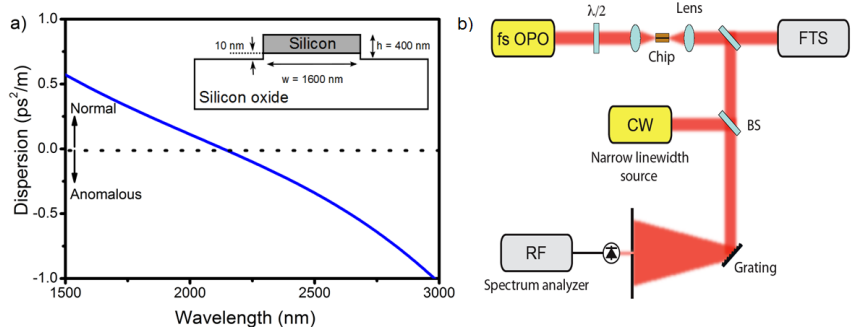

Figure 1: a) The simulated dispersion of the quasi TE-mode of the photonic wire waveguide and b) the experimental setup. a) The zero-dispersion wavelength is at 2,180 nm, while the dispersion is normal at shorter wavelengths and anomalous at longer wavelengths. The waveguide cross-section is shown in the inset. b) experimental setup: the OPO pumped by a Ti-Sapphire mode-locked laser is coupled to the silicon chip with a lens. The output of the chip can be send to a photodetector or a spectrometer.

## 3. Experimental setup and results

The experimental setup for supercontinuum generation is shown in Figure 1 b). The frequency comb seed source consists of a homemade mid-infrared singly resonant optical parametric oscillator (OPO) [13] at a repetition frequency of 100 MHz, synchronously pumped by a femtosecond mode-locked Ti-Sapphire laser. The OPO is tuned to a center wavelength of 2,290 nm, close to the zero dispersion wavelength of 2,180 nm of the silicon waveguide. Pumping a waveguide close to the zero dispersion wavelength in the anomalous region allows for broadband supercontinuum generation [4]. The OPO has a pulse duration of 70 fs, while its average power is 35 mW. The ultra-short mid-infrared fs pulses coming from the OPO are coupled to the quasi-TE mode of the silicon photonic wire using a high NA (NA=0.85) chalcogenide lens with a focal length of 1.87 mm. The output of the chip is coupled, using another chalcogenide lens, to a Fourier transform spectrometer (FTS) to quantify the spectrum of the output pulses. The coupling loss at the input waveguide facet is estimated to be 12 dB, leading to an on-chip peak power of 225 W or pulse-energy of 16 pJ. The spectra at the input and output of the waveguide are shown in Figure 2) for a pulse energy of 16 pJ. The spectrum of the pulses is significantly broadened in the silicon photonic wire waveguide and spans more than an octave: the 30 dB bandwidth spans from 1,540 nm up to 3,200 nm at the output. The peak at 1,600 nm is located in the normal dispersion regime of the waveguide and is generated through dispersive wave generation, a method used to spectrally extend a supercontinuum [14].

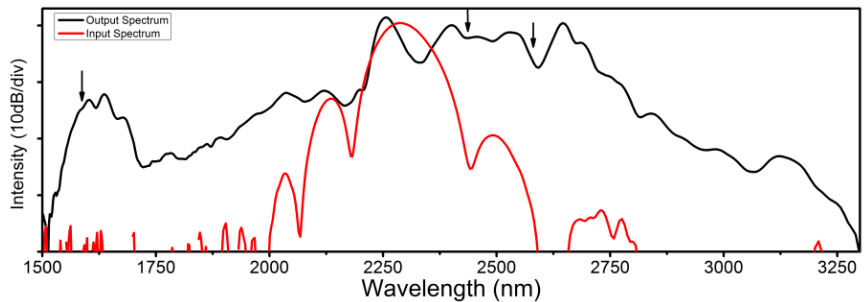

Figure 2: The spectrum at the input (red) and the output (black) of the silicon nanowire. The input pulses are centered at 2,290 nm and have a coupled peak power of 225 W. The pulses are broadened in the silicon photonic wire such that the spectrum of the output pulses spans more than an octave: the 30 dB bandwidth spans from 1,540 nm to 3,200 nm. The arrows indicate the wavelength position where the phase coherence measurements are performed.

We experimentally investigate the phase coherence of the supercontinuum generated in the waveguide using free-running beat note measurements with a set of narrow line-width continuous-wave lasers. First, we beat the free-running seed source with a tunable continuous wave OPO (Argos Aculight, line-width ~60 kHz at 500 µs) at 2,400 nm on a fast InGaAsSb photodetector (Figure 3a)). We then beat the supercontinuum output with the same OPO (Figure 3b) and 3c) respectively), tuned at 2,418 and 2,580 nm. We finally beat (Figure 3 d)), on a fast InGaAs detector, the supercontinuum with a narrow line-width erbium doped fiber laser (Koheras AdjustiK E15, NKT Photonics, line-width 0.1 kHz at 100 µs) at 1,586 nm, far from the seed wavelength. All radio-frequency spectra are recorded with a 100-kHz resolution bandwidth, and a spectrum with a 105-MHz span shows three isolated lines. The strong beat signal at 100 MHz corresponds to the repetition frequency of the fs OPO, while the other two beat notes correspond to the beat signal generated by the continuous wave lasers and the two spectrally closest lines of the frequency comb. The line-width of the beat notes, measured with a 10-kHz resolution bandwidth (inset of the figures) is limited by the instabilities of the free-running lasers but it is found to be about 50 kHz, without noticeable broadening relative to the fs OPO seed source. This demonstrates the frequency comb structure of the supercontinuum.

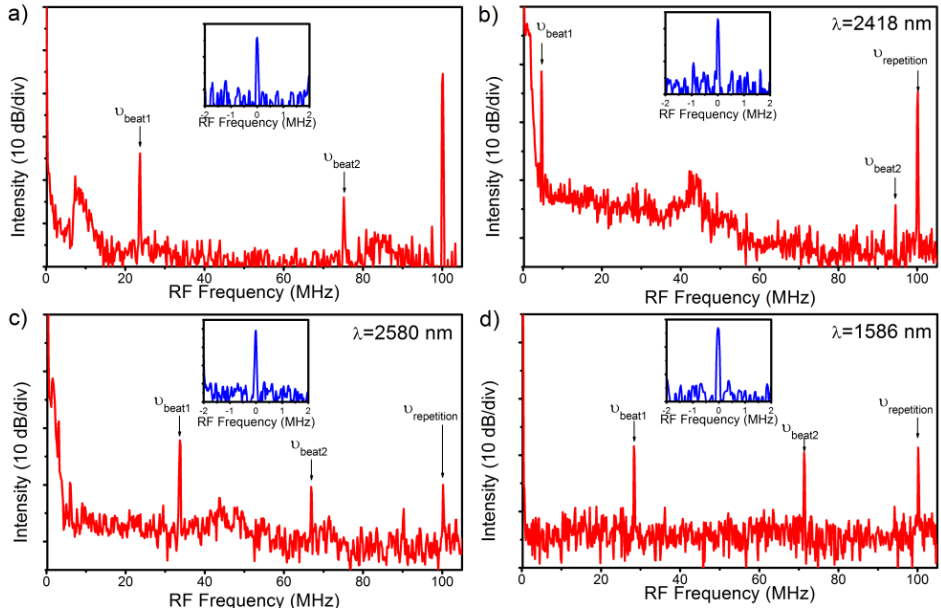

Figure 3: RF spectra showing the narrow line-width beat notes of the input pulses (a) and output pulses (b,c,d). a) RF spectrum of the free-running beat note of the pump pulses and a narrow line-width source at 2,400 nm. b),c),d): free-running beat notes of the spectrally broadened pulses and a narrow line-width source at λ=2,418 nm, λ=2,580 nm and λ=1,586 nm, respectively. The insets in the figure show a high resolution spectrum of the free-running beat notes. The free-running beat notes of the output pulses are measured to be about 50 to 70 kHz wide and are not broadened as compared to beat note measured on the input pulses.

The coherence of the supercontinuum can be simulated and such simulations can be used to confirm the frequency comb structure at the probed wavelengths as well as indicating the coherence over the whole bandwidth. The supercontinuum generation can be simulated by solving the generalized nonlinear Schrödinger equation numerically with a split step Fourier method [4]. The simulation takes the linear propagation loss, the nonlinear phase shift, the three photon absorption and both the induced absorption and dispersion by the carriers into account. In the simulation the nonlinear parameter γ is assumed to be 38 $(Wm)^{-1}$, the linear propagation loss is assumed to be 0.1 $dB.cm^{-1}$ and the three photon absorption coefficient is assumed to be 0.025 cm3.GW$^{-2}$ [15]. Figure 4 a) shows the evolution of the spectrum of a 225-W peak power, 70-fs long pulse as it is propagating along the silicon photonic wire waveguide. The simulated spectrum after 1-cm propagation is shown in Figure 4b). As shown in the figure, the simulation agrees very well with the experimental results. The simulation of the spectral evolution of the pulse along the photonic wire length reveals (Fig. 4 a)) that in the first millimeter of propagation the spectrum is primarily broadened due to self phase modulation. The spectrum is further broadened into the telecom wavelength range, where the group velocity dispersion of the waveguide is normal, through dispersive wave generation [16]. The use of the short pulses favors the processes such as dispersive wave generation and self-phase modulation. Unlike in [17] where longer, ps pulses were used and the spectral broadening primarily results of amplification of background noise (modulation instability) the nonlinear process of dispersive wave generation and self-phase modulation maintain the coherence in the pulse.

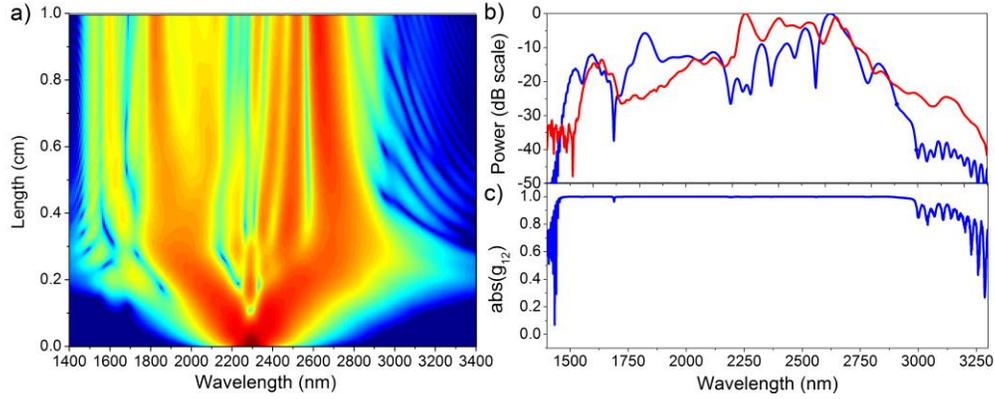

Figure 4: The simulated spectral broadening in the silicon photonic wire waveguide and the coherence of the pulses. a) Evolution of the spectral content of the optical pulse along the length of the waveguide. b) Simulated spectra after 1 cm of propagation in the silicon photonic wire waveguide (blue) and the measured supercontinuum (red). c) Simulated coherence as a function of wavelength.

The coherence of the supercontinuum can be simulated, by including shot noise at the input. The noise Enoise(t) at the input is assumed to be a random variable with a stochastic distribution < Enoise(t) Enoise(t+τ)> = $\frac{h\upsilon}{2}\delta(t)$, with h the Planck constant and $\upsilon$ the frequency of the photons, and analyzing an ensemble of simulated supercontinua [18]. The first order coherence function

$$g_{12}(\omega) = \frac{\left|< E_i(\omega) E_j(\omega) >_{i \neq j}\right|}{\sqrt{<\left|E_i(\omega)\right|^2 >< \left|E_j(\omega)\right|^2 >}}$$

is calculated for an ensemble of 100 spectra and is shown in Figure 4c). The coherence is close to unity over the whole spectrum, indicating that the generated supercontinuum is coherent over its entire bandwidth. To emphasize the comb structure of the supercontinuum spectrum, which results from the pulse-to-pulse coherence, the spectrum of the pulse train at the output of the chip was simulated with a resolution of 10 kHz in a narrow band interval. The spectrum is simulated by first generating a set of pulses including the input shot noise as discussed above. These pulses were stacked together in a pulse train with a repetition frequency of 100 MHz. The Fourier transform was calculated to generate the spectrum of the pulse train. Figure 5 shows the spectrum of a train of 1,000 pulses, calculated in a 500-MHz interval at 1,586 nm. The independent comb lines can clearly be seen. The inset of Figure 5 shows one individual comb line sampled with a resolution of 10 kHz by calculating the spectrum of a pulse train consisting of 10,000 pulses. In the simulations, the width of the comb lines is only limited by the time window used.

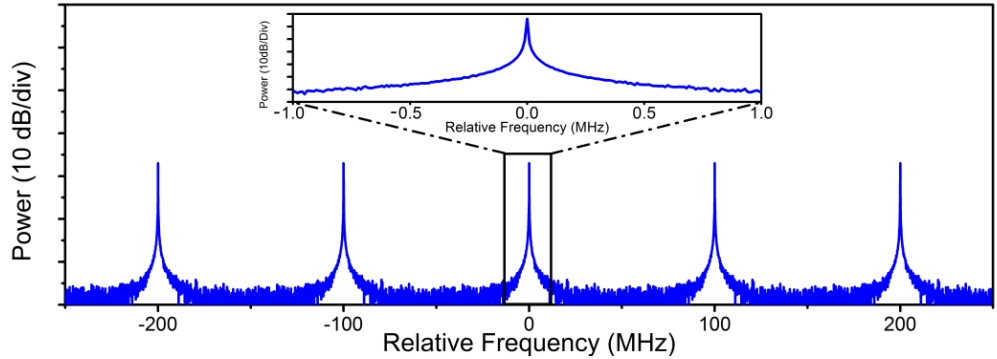

Figure 5: A high resolution spectrum of the broadened output pulses simulated in the vicinity of 1,586 nm (198 THz). The spectrum, simulated over a 500-MHz bandwidth, reveals the comb lines separated by 100MHz in the supercontinuum frequency comb. A high resolution (10 kHz) inset around a comb is also shown.

## 4. Conclusion

Using a silicon nanowire on a chip, we have demonstrated an octave-spanning frequency comb spanning from the telecom wavelength window around 1,500 nm to the mid-infrared wavelength range at 3,300 nm. Improved dispersion engineering could potentially extend the supercontinuum over the whole transparency window of the SOI platform (1,100 nm to 4,000 nm), limited by the buried oxide. Even broader bandwidths could be further obtained up to 5,500 nm with silicon on sapphire waveguide platforms [19, 20]. By using waveguide designs where the buried oxide is removed [21, 22, 23] the entire silicon transparency window (up to 8,500 nm) could be covered. These broadband supercontinua pave the road to self-referenced mid-infrared frequency comb systems, as needed for precision measurements, either in frequency metrology or direct frequency comb spectroscopy, like dual-comb spectroscopy [1].

### Acknowledgements

B. Kuyken acknowledges the special research fund of Ghent University (BOF), for a post doctoral fellowship. We are grateful to Dr. Antonin Poisson and Dr. Clément Lafargue for experimental support. This work was partly carried out in the framework of the Methusalem project "Smart Photonic Chips" and the FP7-ERC-INSPECTRA, FP7-ERC-MIRACLE and FP7-ERC-Multicomb (Advanced Investigator Grant 267854) projects.